# Delay Impact on Stubborn Mining Attack Severity in Imperfect Bitcoin Network

Haoran Zhu, Xiaolin Chang, Jelena Mišić, Vojislav B. Mišić

*Abstract*—Stubborn mining attack greatly downgrades Bitcoin throughput and also benefits malicious miners (attackers). This paper aims to quantify the impact of block receiving delay on stubborn mining attack severity in imperfect Bitcoin networks. We develop an analytic model and derive formulas of both relative revenue and system throughput, which are applied to study attack severity. Experiment results validate our analysis method and show that imperfect networks favor attackers. The quantitative analysis offers useful insight into stubborn mining attack and then helps the development of countermeasures.

*Index Terms*—Bitcoin, Stubborn Mining Attack, Proof-of-Work, Quantitative Analysis.

## I. INTRODUCTION

Stubborn mining attack [1] is a variant of selfish mining in Bitcoin blockchains. Both selfish mining and stubborn mining attack have attracted the attention of many researchers [2]-[7]. It is necessary to study the severity of stubborn mining attack to quantify the impact on Bitcoin system.

Most of the existing works only studied selfish mining and stubborn mining in **perfect** networks where miners receive new blocks at once [2]-[4]. In reality, Bitcoin network is **imperfect** that contains network attacks and/or transmission caused congestion. It takes a period of time for the whole network receiving new blocks and we call this time interval **block receiving delay**. A miner keeps mining until it receives new blocks. As a result, it is possible for a miner to generate another valid block. More than one block is generated at the same height and an **unintentional fork** occurs. Unless otherwise stated, **fork** refers to unintentional fork and **delay** refers to block receiving delay. Malicious miners deviate from honest mining and broadcast blocks strategically, which extends the delay, can also make blockchain fork (called intentional fork).

There are several factors having impact on unintentional fork probability, such as block generation time, bandwidth and other network issues [9]. Among which block propagation delay is one of the main reasons [10]. The intentional fork probability is closely related to blockchain attack itself. Moreover, attackers aim to get profit by making intentional forks rather than creating intentional forks. An intentional fork is a consequence of blockchain attack and thus the analysis on intentional fork probability is meaningless. The existing works show that unintentional fork is extremely beneficial to attackers [6]-[8]. Although the fork probability is low in Bitcoin, there is still a nonnegligible gap between perfect and imperfect networks.

Stubborn mining has 3 basic strategies proposed in [1], *L*ead-stubborn (*L*), equal-*F*ork-stubborn (*F*), and *T*rail-*j*-stubborn (*$T_j$*) strategy. $T_1$ is the dominant strategy in $T_j$ strategy [1] so we omit the scenarios that $j > 1$. Combining the 3 basic strategies, there are also 4 hybrid strategies, Lead-Fork (*LF*), Lead-Trail (*LT*), Fork-Trail (*FT*), and Lead-Fork-Trail (*LFT*). This paper considers these 7 strategies that are detailed in Section II.A.

Since stubborn mining attack was proposed, numerous studies have been focusing on its impact on blockchain systems. Grunspan et al. [3] studied *L* and *F* strategies by state machine and derived formulas of relative revenue. The authors in [4] evaluated stubborn mining attack based on Markov chain in terms of relative revenue, throughput, and other metrics. These works are in perfect network scenario, which may divorce from reality. Wang et al. [5] modeled selfish mining, *T* and *FT* stubborn in imperfect PoW Ethereum. However, their definition of "imperfect" is different from ours. Miners select blocks by timestamp in their system. In other words, they implicitly assume that most of miners are synchronous and few miners have network issues. We consider a more realistic imperfect network and miners select blocks by receiving time. The chain selection protocol of Bitcoin is the longest-chain and GHOST is another widely used protocol. The authors in [8] explored stubborn mining severity in imperfect GHOST Bitcoin. Their model and formulas are hard to evaluate imperfect longest-chain Bitcoin, which is the scenario of this paper. Both their work and this paper can give insight into stubborn mining attack. In summary, the existing works either consider perfect network or some of stubborn mining strategies in imperfect Bitcoin. Moreover, the existing models are hard, if not impossible, to be used to analyze stubborn mining attack in our scenarios.

There are at least the following two challenges to system modeling and formula derivation.

**Challenge 1.** The behaviors of malicious and honest miners are interrelated. Malicious miners behave differently in different stubborn strategies. When a fork occurs, the location of new blocks leads to different blockchain structures. How to capture all the interactions in one model is the first challenge.

**Challenge 2.** Selfish mining attacker is rewarded for sure in some situations. However, these situations are not fully established in stubborn mining. Moreover, possible block generation scenarios are much more in an imperfect network than in a perfect network. How to compute the revenue of stubborn attacker in an imperfect network is a huge challenge.

The above discussions motivate the work of this paper. This paper aims to quantitatively analyze stubborn mining attack severity in imperfect Bitcoin blockchains. The contributions of this paper are summarized as follows.

- We propose an analytical model based on continuous time Markov chain to capture the evolution and dynamics of

imperfect Bitcoin network. Our model can be applied both Bitcoin and most Bitcoin-like blockchain systems.
- We derive formulas for computing miner revenue of stubborn mining attack, and then derive formulas for relative revenue and system throughput which enables the evaluation of stubborn mining severity. Our best knowledge indicates that we are the first to derive the formulas for these metrics of all stubborn mining attack strategies in imperfect Bitcoin.

We make a simulator to cross-validate our model and formulas. The quantitative analysis provides insight into stubborn mining attack and then helps design countermeasures. The remainder of this paper is organized as follows. Section II gives system description, model and metric formulas. Section III and IV present evaluation results under different scenarios and the paper conclusion, respectively.

## II. SYSTEM DESCRIPTION, MODEL AND METRIC FORMULAS

This section first gives preliminary and system description in Section II.A and II.B, respectively. The proposed model and metric formulas are presented in Section II.C and II.D, respectively.

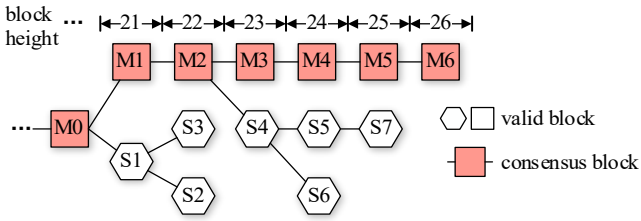

**Fig.1.** Imperfect Bitcoin system

### A. Preliminary

**Bitcoin system.** Bitcoin is a Proof-of-Work (PoW) blockchain system where miners generate blocks by solving hash-based puzzles. Once a puzzle is solved, a **valid block** is generated. The time between two adjacent blocks is called **block creation interval**, about 10 minutes. A valid block is called **consensus block** after acknowledged by all miners. The miner of consensus block gets block generation revenue, which is halved every 210,000 consensus blocks. Although miners also get transaction fee, it is far less than block generation revenue. Thus, this paper only considers block generation **revenue** and normalizes it to 1 per consensus block. **Height** denotes the number of blocks from the block to **genesis block** (first block of blockchain) and **distance** is the height difference. The chain from the genesis block to the highest consensus block is **main chain**. Blocks other than consensus blocks on main chain are **stale block**s. The chain from the highest consensus block to **leaf block** (block with maximum height) is named as a branch and the number of blocks is branch **length**. In Fig.1, assume there are already 20 consensus blocks, then the height of "S1" is 21. The distance between "M1" and "S4" is 2. "M6" is a leaf block. "S1"~"S7" are stale blocks.

**Pools and mining.** Miners get nothing until they produce consensus blocks. They usually form mining **pool**s to combine their computing power for mining. A pool has a manager determining the block on which miners mine and publishing valid blocks. A miner cannot mine on other blocks otherwise it will be detected as indiscipline and removed. All miners share the revenue equally. Once a miner owns a consensus block, all miners in the pool get revenue and the total revenue of all miners is 1 (6.25 BTC in reality). A pool mines block **honestly** by default, which can be described detailly as follows.
- Pools mine blocks continuously on the longest branch.
- Pools broadcast blocks as soon as they find/receive them.
- Pools may create unintentional forks caused by delay. At least one block is generated in a block generation interval.
- When there is more than one branch with the same length, pools mine on the one that contains their own block if exists. Otherwise, they select a branch randomly.

**Stubborn mining strategies.** Let $\Delta$ denote length difference between private and public branches. Compared with selfish mining, the basic strategies can be described as follows.
- $L$ strategy: When $\Delta = 2$ and honest miners produce a new block, attackers only publish one unpublished block.
- $F$ strategy: When both private and public branches have the same positive length (denoted as $\Delta = 0'$) and attackers produce a new block, they do not publish this block.
- $T_j$ strategy: When private branch falls $j$ block(s) behind public branch, the attackers still mine on private branch, denoted as $T_j$. We only study the dominant strategy $T_1$ [1]. When $\Delta = -1$ and attackers generate a new block, they publish the block. The length of public and private branches is equal but all other miners mine on public branch. $\Delta = 0''$ denotes this case.

Selfish mining and 3 basic stubborn mining strategies in imperfect Bitcoin are summarized in TABLE I. The behaviors of hybrid strategies are the combination of the different behaviors of basic strategies (italics in TABLE I).

TABLE I
STUBBORN AND SELFISH MINING IN IMPERFECT BITCOIN

| | State | S | L | F | T |
|---|---|---|---|---|---|
| **Attacker finds a block** | $\Delta \geq 0$ | hold | | | |
| | $\Delta = 0'$ | publish | | *hold* | publish |
| | $\Delta = 0''$ | -- | | | *publish* |
| | $\Delta = -1$ | -- | | | *publish* |
| **Honest miners find $i$ block(s) ($i>0$)** | $\Delta \geq 1$ and $\Delta \neq 2$ | publish 1 | | | |
| | $\Delta = 2$ | publish all | *publish* 1 | publish all | |
| | $\Delta = 0'$ | mine on new blocks in pri if exist, otherwise mine pb* | | | otherwise *mine pri* |
| | $\Delta = 0''$ | -- | | | *mine on pri* |
| | $\Delta \leq 0$ | mine on one of the new blocks | | | |

\* "pb" denotes public branch, "pri" denotes private branch.

### B. System Description

There are two types of miners, malicious miner and honest miner. All malicious miners conspire and form a malicious pool (**MP**). Other miners (namely, honest miners) form several honest pools (**HP**s). The MP conducts stubborn mining attack and HPs mine blocks honestly. The MP can quickly detect new blocks from other pools and propagate its block through the assistance of eclipse attack and network sniffing [1]. We assume that miners in the MP are well-connected and MP does not make unintentional forks. We call the branch that MP mines on as **private branch** and other branch(es) is **public branch**(es).

Computing power is a key property of a pool, which is proportional to the probability that a pool generates blocks. The computing power of the largest pool in Bitcoin is around 25%~35% [12]. A pool with more than a half computing power can conduct double-spending attack and get all revenue, which is meaningless to study. Furthermore, the existing works show that small pools conducting stubborn mining attack get no benefits [1][4]. To make our work exhaustive and meaningful, we study the scenarios that the computing power of MP varies from 5% to 45%. The results can show the profit of the stubborn mining attack to large pools and the loss to small pools. There are several assumptions in the system we consider.

- The total computing power is constant and all pools keep mining. A miner does not join or quit its pool temporarily.
- There is only one group of attackers. Namely, the number of MP is one. They mine blocks by stubborn mining strategy.
- No more than two blocks can be created in a block generate interval since the small probability of three-chain fork [13]. That is, there are two new blocks when a fork occurs.

TABLE II
DEFINITION OF NOTATIONS

| Notation | Definition |
|---|---|
| $\Delta$ | The length difference between the private and public branches. |
| $N$ | The number of leaf block(s) where HPs can mine. |
| $\theta$ | Probability of unintentional fork occurrence. |
| $\alpha, \beta$ | The total computing power of MP and HPs, respectively. They also represent the rates of MP and HPs generate block(s). |
| $\beta_1, \beta_2$ | Rates of HPs generating a block and a fork in a block generating interval, respectively. $\beta=\beta_1+2\beta_2$, $\beta_2=\beta\theta$. |
| $P_{\beta 1}$ | Probability of HPs generating one block. $P_{\beta 1}=\beta(1-\theta)$. |
| $\gamma_N$ | Probability that HPs generate one block on private branch when there are $N$ leaf blocks. |
| $g_N^B$ | Probability that HPs generate a fork on $B^*$ when there are $N$ leaf blocks. |
| $P_H(x)$ | Probability that a public branch becomes the main chain in Scenario x. It is specified in Section II.D. |

\* $B \in \{A, AH, H\}$ denotes the fork scenarios. $A$, $AH$ and $H$ denote HPs generate two blocks behind attacker's branch (private branch), attacker's and one honest branch (public branch), honest branch(es), respectively.

### C. System Model

It is assumed that the block generation follows exponential distribution as in [5]-[8]. We design a continuous time Markov chain to describe the Bitcoin dynamics in an imperfect network. System state is defined as a two-tuple ($\Delta$, $N$). $\Delta$ denotes length difference between private and public branches as in Section II.C. $N$ denotes the number of blocks that pools can mine on. Fig.2 describes the state-transition diagram. TABLE II gives the symbol definition.

There are some constraints on the tuples. Since the existence of genesis block, at least one leaf block exists, $N \geq 1$. MP only produce blocks on private branch and does not produce fork. Thus, there is only one private branch. On the contrary, HPs can generate more than one public branch due to being in imperfect network. HPs mine on any branch, including public branch and published private branch. Recall that two new blocks are generated when a fork occurs, $3 \geq N$. With these descriptions and assumptions in Section II.B, $\Delta \in \{-1, 0'', 0'\} \cup \{0, 1, 2, \ldots\}$ and $N \in \{1, 2, 3\}$.

Some lines are split in two and one of them is marked with a word in Fig.2. For example, the state (2,3) transits to (0,1) with a green line and transits to (1,2) with a green line with a mark "$L$". This denotes that if MP applies $L^*$ strategy ($L$, $LF$, $LT$, $LFT$), state (2,3) transits to (1,3) at a rate of $\beta_1$. Otherwise, (2,3) transits to (0,1) at the same rate.

### D. Metric Formulas

Let $\pi(\Delta, N)$ denote the steady-state probability of state ($\Delta$, $N$). The steady-state probabilities of all states can be obtained by solving global balance equations and they are applied to calculate system metrics.

***Revenue of pools.*** The revenue of pools is a weighted average of the expected revenue in each system state. The weight is the steady-state probability. By considering all block generation events, we can get expect revenue in each state and then get the revenue of pools. However, it is challenging to calculate revenue of pools in imperfect network directly. We first compute $P_H(len)$ (probability that public branch becomes main chain when $\Delta = len$), $P_H(0', N)$ (probability that public branch becomes main chain in state $(0', N)$) in ***Proposition* 1** and $P_F$ (probability that MP's new-finding block becomes consensus block in state $(0', N)$) in ***Proposition* 2**. Pools' revenue is given

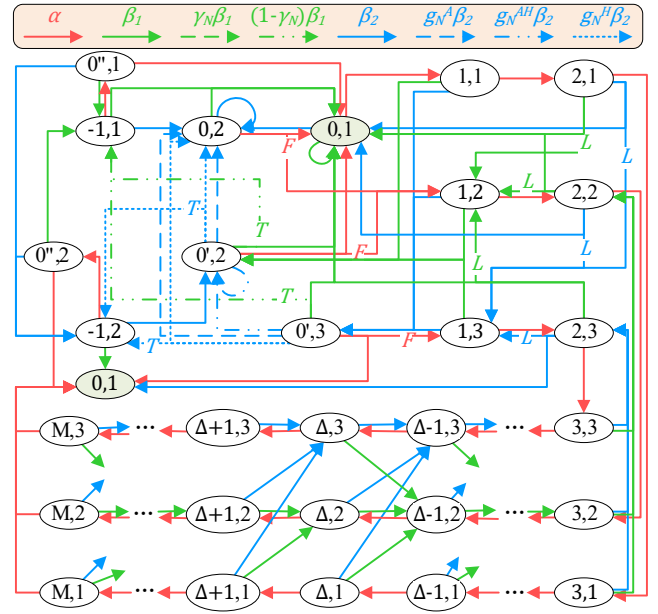

**Fig.2.** State-transition diagram

in ***Proposition* 3**. Due to space limitation, we omit the proof of these three propositions.

***Proposition* 1.** $P_H(0', N)$ and $P_H(len)$ can be computed by (1).

$$\begin{cases} P_H(-1) = \beta/(1-\alpha\beta) \\ P_H(0'') = \beta^2/(1-\alpha\beta) \\ P_H(0',N) = P_{\beta 1}(1-\gamma_N)((1-T)+P_H(-1)T) \\ \quad + \beta_2 g_N^{AH} P_H(0',2) + \beta_2 g_N^H ((1-T)+P_H(-1)T) \\ \quad + \alpha\left(PH_N + \alpha P_{\beta 1}(1-\gamma_N)PH_2 + \beta_2(g_N^H + 0.5 g_N^{AH})PH_3 L\right)F \end{cases} \quad (1)$$

where $PH_N = P_{\beta 1}(1-\gamma_N)P_H(0',2) + \beta_2(g_N^H + 0.5g_N^{AH})P_H(0',3)$, and $L$, $F$, $T$ are flags of MP's strategy.

**Proposition 2.** $P_F$ can be calculated by (2).

$$\begin{aligned} P_F &= P_{\beta 1}(1 - P_H(0',2)) + \beta_2(1 - P_H(0',3)) \\ &+ \alpha L \begin{pmatrix} 1 - \beta\left(P_{\beta 1}P_H(0',2) + \beta_2 P_H(0',3)\right) \\ + P_{\beta 1}\left(P_{\beta 1}\gamma_2 P_H(0',2) + \beta_2\left(0.5 g_2^{AH} + g_2^A\right)P_H(0',3)\right) \\ + \beta_2\left(P_{\beta 1}\gamma_3 P_H(0',2) + \beta_2\left(0.5 g_3^{AH} + g_3^A\right)P_H(0',3)\right) \end{pmatrix} \end{aligned} \quad (2)$$

**Proposition 3.** Let $P_H(0') = P_{\beta 1}P_H(0',2) + \beta_2 P_H(0',3)$. Revenue of MP and HPs can be computed by (3).

$$\begin{cases} E_M = \pi(0,1)\alpha\left(\alpha\left((1-L) + (1-\beta^2 P_H(0'))L\right)\right) + \pi(0,1)\alpha \cdot \\ \quad \beta_{i-1}\alpha((1-F)+P_F F) + \beta_{i-1}P_{\beta 1}(\gamma_i + (1-\gamma_i)(1-P_H(-1))T) \\ \quad + \beta_{i-1}\beta_2\left(g_i^A + g_i^{AH} P_F/(P_{\beta 1}(1-\gamma_i)) + g_i^H(1-P_H(-1))T\right) \\ \quad + (\pi(0,2)+\pi(0',N))\alpha((1-F)+P_F F) \\ \quad + \pi(\Delta,N)\alpha\left((1-L) + (1-\beta^\Delta P_H(0'))L\right) \\ \quad + \pi(0'',N)\alpha T + \pi(-1,N)T\alpha^2/(1-\alpha\beta) \\ E_H = \pi(0',N)\beta_1\left(\gamma_N + (1-\gamma_N)((1-T)+P_H(-1)T)\right) \\ \quad + \pi(0',N)\beta_2\left(g_N^A + g_N^{AH} + g_N^H((1-T)+P_H(-1)T)\right) + \\ \quad \pi(1,N)\beta_{i-1}\left(\alpha F(1-P_F) + P_{\beta 1}(1-\gamma_i)((1-T)+P_H(-1)T)\right) \\ \quad + \pi(1,N)\beta_{i-1}\beta_2 g_i^H((1-T)+P_H(-1)T) + \pi(1,N)\beta_{i-1} \cdot \\ \quad \beta_2 g_i^{AH}\left((1-P_F/P_{\beta 1}(1-\gamma_i))F + P_H(0',2)(1-F)\right) \\ \quad + (\pi(0,1)+\pi(0,2)+\pi(-1,N))(\beta_1+\beta_2) + \pi(\Delta,N) \cdot \\ \quad (\beta_1+\beta_2)\beta^\Delta P_H(0')L + \pi(0'',N)(\beta_1+\beta_2)P_H(-1)T \end{cases} \quad (3)$$

$i \in \{2,3\}, \Delta \geq 1$

*Relative revenue of pools.* The total revenue of all pools is varying with attack happening. That is, the increase in revenue of MP may not mean the decrease in revenue of HPs. Thus, we use relative revenue, the proportion of pool's revenue in total revenue, to measure the proceeds from launching stubborn mining attack. A gain in MP's relative revenue ($RR_M$) means the loss of HPs' relative revenue ($RR_H$).

*Transactions per second (TPS).* We use TPS to evaluate the system throughput and only the transactions in consensus blocks are accounted. The number of transactions in most blocks is around 1000-3000 in Bitcoin [12]. To get a more general result, we normalize transaction number in a block to 1 and thus TPS can be evaluated by consensus block generation rate. As a consensus block produces one unit revenue, consensus block generation rate is equal to total revenue.

### III. EXPERIMENT AND RESULTS

The computing power of MP varies from 5% (0.05) to 45% (0.45). It is hard for a pool to detect delay because of the asynchronous network. It is reasonable to use fork probability to denote study the delay impact. The reasons are as follows. The data in [10] show that a higher delay results in a higher fork probability. Our investigation of recent blocks in real Bitcoin indicates that the fork probability is around 1% [11]. There are also Bitcoin variants which produce blocks in seconds (such as Bitcoin Fast) and delay has a higher impact on fork probability. They have a higher fork probability at around 5%. To investigate Bitcoin system comprehensively, we set the fork probability $\theta$ to 1% (low probability), 5%, 10% (high probability), and 20% (super high probability).

Only the work in [6] (denoted as Yang's model) is similar to ours. They considered selfish mining attack in imperfect Bitcoin system, which can be seen as a special case of ours by setting $L$, $F$, $T$ = 0, 0, 0. We reproduce their experiments to validate our model and formulas from one aspect. We also develop a simulator by C language to verify our work in general scenarios. In each round of the simulation, a blockchain is created with 10 million consensus blocks and we do 300 rounds for each group of parameters. Other parameters settings are shown in TABLE III. The numerical experiments are conducted in MAPLE [14].

TABLE III
PARAMETERS SETTINGS

| Notation | Setting | Notation | Setting |
|---|---|---|---|
| $\gamma_N$ | $1/N$ | $g_N^A$ | $\gamma_N^2$ |
| $g_N^{AH}$ | $2\gamma_N(1-\gamma_N)$ | $g_N^H$ | $(1-\gamma_N)^2$ |

From the experiment results, we observe that
- Our model and formulas can capture the system dynamics and compute metrics accurately (shown in Fig.3).
- *LFT* strategy is the most profitable strategy for large pools (more than 35% computing power) but is the dominated strategy for small pools (less than 35% computing power) (shown in Fig.4).
- The key to get a higher relative revenue is an increase in computing power. The existence of unintentional forks benefits MP to conduct stubborn mining attack and does harm to HPs (shown in Fig.5).
- *LFT* strategy benefits large pools that have more than 30% total computing power (shown in Fig.5). Small pools should give up stubborn mining strategies.
- The benefit threshold decreases with the increasement of fork probability (shown in Fig.5). Attackers have the motivation to launch network attacks to increase fork probability.
- Stubborn mining attack downgrades TPS greatly, especially for attacker with large computing power (shown in Fig.6).

Compared with small attackers, fork probability has less influence on TPS for large attackers.

## IV. CONCLUSION

In this paper, the severity of stubborn mining strategies is investigated by developing both a Markov model and metric formulas. Experiment results validate that (i) stubborn mining attack is very profitable for attackers with large computing power, (ii) the attack extremely downgrade the system throughput, and (iii) high fork probability benefits attacker.

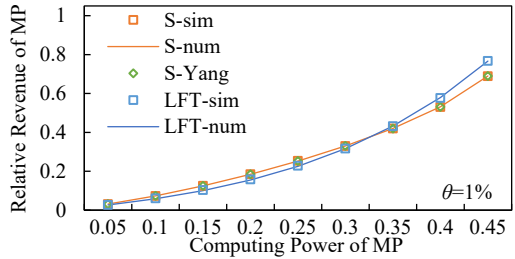

**Fig.3.** Verify our model and formulas under simulation and Yang's model.

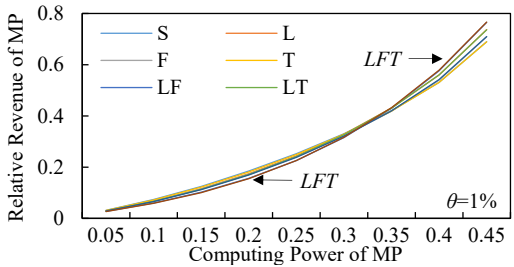

**Fig.4.** Relative revenue of MP under selfish mining and each stubborn mining strategies when $\theta=1\%$.

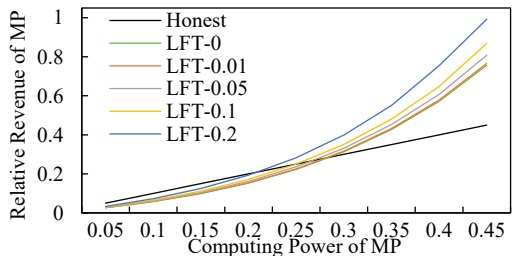

**Fig.5.** Relative revenue of MP under *LFT* strategy in each fork probability over computing power of MP.

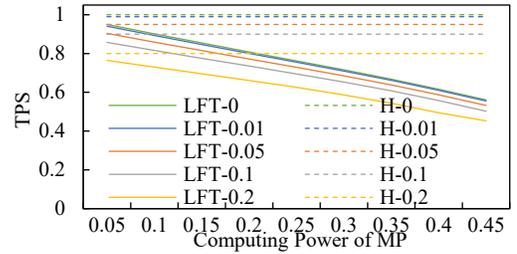

**Fig.6.** TPS under *LFT* strategy in each fork probability over computing power of MP